\documentclass[10pt]{article}
\usepackage{amsmath}
\usepackage{amsfonts}
\usepackage{amssymb}
\usepackage{bm,bbm}

\usepackage{epsf,graphicx,rotating,color}

\newcommand{\rmd}{{\rm d}}
\newcommand{\rme}{{\rm e}}
\newcommand{\rmi}{{\rm i}}

\newcommand{\la}{\langle}
\newcommand{\ra}{\rangle}
\newcommand{\lla}{\left\langle}
\newcommand{\rra}{\right\rangle}

\newcommand{\eps}{\varepsilon}
\newcommand{\Sp}{{\it Sp}}
\newcommand{\Id}{\boldsymbol{1}}
\newcommand{\cM}{{\cal M}}
\newcommand{\cN}{{\cal N}}

\newcommand{\ts}{\textstyle}

\newcommand{\One}{\mathbbm{1}}

\title{Monomial integrals on the classical groups}

\author{\small T. Gorin and G. V. L\' opez\\
    \small Departamento de F\'\i sica, Universidad de Guadalajara\\
    \small Blvd. Marcelino Garc\'\i a Barragan y Calzada Ol\'\i mpica\\
    \small  44840 Guadalajara, Jalisco, M\' exico}

\begin{document}

\maketitle

\footnotetext{gorin@pks.mpg.de}

\begin{abstract}
This paper presents a powerfull method to integrate general monomials on the
classical groups with respect to their invariant (Haar) measure. The method 
has first been applied to the orthogonal group in [J. Math. Phys. 43, 3342 
(2002)], and is here used to obtain similar integration formulas for the 
unitary and the unitary symplectic group. The integration formulas turn out to 
be of similar form. They are all recursive, where the recursion parameter is 
the number of column (row) vectors from which the elements in the monomial are 
taken. This is an important difference to other integration methods. The 
integration formulas are easily implemented in a computer algebra environment, 
which allows to obtain analytical expressions very efficiently. Those 
expressions contain the matrix dimension as a free parameter.
\end{abstract}



\section{Introduction \label{I}}

With the classical groups, we mean the orthogonal group $O(d)$, the unitary 
group $U(d)$, and the unitary symplectic group $\Sp(2d)$~\cite{Weyl39}. They
all possess a unique invariant measure, the ``Haar'' measure~\cite{Haar33}, 
which is the integration measure commonly used.

Integration formulas for the classical groups are of interest in various fields
of mathematical physics. A number of different classes of integrals have been 
studied, among those: generating functions, such as the 
Harish-Chandra-Izykson-Zuber integral~\cite{HarCha57,IZ80}, 
and monomial integrals, which are the concern of the present work. 
In some cases, the corresponding integral can be solved by character
expansion due to Balantekin~\cite{Bal84} and Balantekin and 
Cassak~\cite{Bal02}. Integration formulas for certain simple monomials 
have been developed by nuclear physicists~\cite{UllPor63,Ullah64}. This work 
was motivated by the fact that 
statistical methods based on the classical groups are very successful in 
describing certain aspects of nuclear reactions (see~\cite{Porter67,Bro81} and 
references therein). Later, Mello and Seligman devised an algebraic method to 
compute low order monomial integrals on $U(d)$~\cite{MelSel80}, and Samuel 
solved the problem in full generality~\cite{Sam80}. The result was an explicite 
formula for arbitrary monomial integrals over $U(d)$.
The method in~\cite{Sam80} is based on the 
representation theory of $U(d)$, and we call it the {\em group theoretical
method}. Some 25 years ago, it has become clear that group integrals on the
classical groups play an important role in mesoscopic 
transport~\cite{Mel90,BroBee96,MelKum04}, in quantum chaos~\cite{Haake,Mehta}, 
and more recently in some aspects of quantum information and decoherence, 
{\it e.g.}~\cite{GS02,GS03,Valle07}. 

These applications lead to a renewed interest in efficient methods for the
analytical calculation of monomial integrals. An interesting unconventional 
approach has been devised by Prosen et al.~\cite{PSW02}, and very
recently, an improved {\em invariant method} has been developed for monomials 
in $U(d)$~\cite{Aub03,Aub04}, and also in $O(d)$~\cite{Braun06}. Finally,
Collins and {\' S}niady presented a group theoretical approach, which allows
to compute monomial integrals over all three classical groups~\cite{ColSni06}.

In~\cite{Gor02}, a further method has been developped, which is very different
from the previous ones; the {\em column vector 
method} as it might be called. It lead to a new type of integration formula 
for monomial integrals on $O(d)$. This method is easily implemented in 
a computer algebra language. It allows to compute arbitrary monomial integrals 
very efficiently, where the result will always be a rational function in the
matrix dimension of the group. The purpose of the present paper is to 
apply that method to $U(d)$ and $\Sp(2d)$. In section~\ref{G} the matrix
representations of the classical groups are introduced, and the notation for 
the different group integrals used is defined. The general idea of the column 
vector method, as well as the result for $O(d)$ is reviewed in section~\ref{O}. 
In the sections~\ref{U} and~\ref{S} the integration formulas for the unitary
and the unitary symplectic group are derived. A summary is given in
section~\ref{C}.

\section{\label{G} General considerations}

In this section, we introduce the fundamental matrix representations of the 
groups under consideration. Then, we define the integration measures and 
specify some notational conventions related to the integrals to be calculated.

\subsection{\label{GO} The orthogonal group \boldmath $O(d)$}
An element $w\in O(d)$ is a $d$ dimensional square matrix with real entries 
$(w_{ij})$, where $1\le i,j\le d$. In addition, $w$ fullfills the following
orthogonality relations:
\begin{equation}
w\; w^T = w^T\; w = \Id \quad\Leftrightarrow\quad 
\forall\; 1\le i,j\le d\; :\; \sum_{k=1}^N w_{ik}\, w_{jk} = \delta_{ij}\; .
\end{equation}
The most general monomial of matrix elements of $w$ is denoted by $\cM_1(w)$.
The subscript $1$ will be replaced by $\beta=2(4)$ in the unitary
(unitary symplectic) case. Here, $\beta$ is the symmetry parameter of the
respective group.
\begin{equation}
\cM_1(w)= \prod_{r=1}^p w_{I_r J_r} = \prod_{i,j=1}^d w_{ij}^{M_{ij}} \qquad
M_{ij}= \sum_{r=1}^q \delta_{i I_r}\, \delta_{j J_r} \; ,
\label{GO:defmon}\end{equation}
where $q$ is the order of the monomial, and $I$ and $J$ are multi-indeces of 
dimension $q$. The notation with the multi-indeces $I,J$ is the most common 
one, but in the present approach the matrix notation is more convenient. 

For the integration on $O(d)$, the normalized Haar measure is used. For the
classical groups, in general, it will be denoted by $\sigma_\beta$, where 
$\beta$ is the symmetry parameter introduced above. In the present case, the 
integral of an arbitrary monomial is denoted by
\begin{equation}
\la M\ra = \int\rmd\sigma_1(w)\; \cM_1(w) \; .
\label{GO:defint}\end{equation}
The normalization is such that $\la o\ra = 1$, where $o$ is the matrix with 
zero elements everywhere.

The column vector method has been applied to the orthogonal group,
first~\cite{Gor02}. The results obtained therein are reviewed in
section~\ref{O}. It allows us to introduce the general idea
of the method, as well as a number of notational conventions. The latter will
be useful for the unitary and unitary symplectic case, also.

\subsection{\label{GU} The unitary group \boldmath $U(d)$}

An element $w\in U(d)$ is a $d$ dimensional square matrix with complex entries
$(w_{ij})$, where $1\le i,j\le d$. In addition, $w$ fullfills the following
orthogonality conditions: 
\begin{equation}
w\; w^\dagger = w^\dagger\; w = \Id \quad\Leftrightarrow\quad 
\forall\; 1\le i,j\le d\; :\; \sum_{k=1}^d w_{ik}\, w_{jk}^* = \delta_{ij}\; .
\end{equation}
The most general monomial on the unitary group depends on matrix elements of 
$w$ and $w^\dagger$. It will be denoted by $\cM_2(w)$:
\begin{equation}
\cM_2(w)= \prod_{r=1}^p w_{I_r J_r} \prod_{s=1}^q w^*_{I'_s J'_s} =
  \prod_{i,j=1}^d w_{ij}^{M_{ij}}\; (w^*_{ij})^{N_{ij}} \qquad
M_{ij}= \sum_{r=1}^p \delta_{i I_r}\; \delta_{j J_r}\qquad
N_{ij}= \sum_{s=1}^q \delta_{i I'_r}\; \delta_{j J'_r} \; ,
\label{GU:defmon}\end{equation}
where $I,J$ and $I',J'$ are multi-indeces of dimension $p$ and $q$,
respectively. Again, while the notation with the multi-indeces $I,J$ and 
$I',J'$ is the most common one, the matrix notation with $M$ and $N$ is more
convenient for our purpose.

For $U(d)$, the normalized Haar measure is $\sigma_2(w)$, while the
monomial integral is denoted by
\begin{equation}
\la N|M\ra = \int\rmd\sigma_2(w)\; \cM_2(w) \; .
\label{GU:defint}\end{equation}
The integration formula for the unitary group is derived in section~\ref{U}.

\subsection{\label{GS} The unitary symplectic group \boldmath $\Sp(2d)$}

The unitary symplectic group may be defined as the subgroup of $U(2d)$ which
is invariant under the antisymmetric billinear form
\begin{equation}
Z'= \begin{pmatrix} o & \One_d\\ -\One_d & o\end{pmatrix} \quad :\quad
\Sp(2d)= \left\{ \bm w\in U(2d)\; |\; 
   \bm w^T Z' \bm w = Z' \right\} \; .
\end{equation}
In order to fulfill this invariance condition, the matrices $\bm w$ must be of
the form
\begin{equation}
\bm w = \begin{pmatrix} z^* & w \\ -w^* & z\end{pmatrix} \; ,
\end{equation}
whith complex $d$-dimensional square matrices $w$ and $z$. For such matrices, 
the unitarity conditions become:
\begin{equation}
\la\vec w_\mu|\vec w_\nu\ra + \la\vec z_\mu|\vec z_\nu\ra = \delta_{\mu,\nu}
\qquad
\la\vec z^*_\mu|\vec w_\nu\ra - \la\vec w^*_\mu|\vec z_\mu\ra = 0 \; ,
\label{GS:ONrels}\end{equation}
where $\vec w_\mu$ and $\vec z_\mu$ denote the respective column vectors of 
the matrices $w$ and $z$. This parametrization will be used to perform the 
integration over $\Sp(2d)$. At this point we may note that
\begin{equation}
\forall \bm w\in\Sp(2d) \; : \; \bm w^{-1}= \bm w^\dagger = \begin{pmatrix}
   z^T & -w ^T \\ w^\dagger & z^\dagger\end{pmatrix} \quad\text{such that}\quad
   \bm w^\dagger \bm w = \bm w \bm w^\dagger = \begin{pmatrix} \One_d & o\\
   o & \One_d\end{pmatrix} \; ,
\label{GS:inverse}\end{equation}
where $\bm w^\dagger$ is again in $\Sp(2d)$.
The most general monomial on the unitary symplectic group contains matrix 
elements from four matrices: 
$w ,\, w^\dagger ,\, z$, and $z^\dagger$. It is denoted by:
\begin{equation}
\cM_4(\bm w)= \cM_4(w,z)= \prod_{i,j=1}^d w_{ij}^{M_{ij}}\; 
   z_{ij}^{M'_{ij}}\; (w^*_{ij})^{N_{ij}}\; (z^*_{ij})^{N'_{ij}} \; ,
\label{GS:defmon}\end{equation}
where $M, M', N, N'$ are $d$-dimensional square matrices with
non-negative integer entries. If we collect these matrices in an appropriate
way in the $2$$\times$$2$-matrix, we may equally well write:
\begin{equation}
\bm M = \begin{pmatrix} N' & M\\ N & M'\end{pmatrix} \quad :\qquad
\cM_4(\bm w)= (-)^{\bar N}\prod_{i,j=1}^{2d} \bm w_{ij}^{\bm M_{ij}} \; ,
\qquad \bar N= \sum_{i,j=1}^d N_{ij} \; .
\end{equation}
For ${\it Sp}(2d)$, we denote the normalized Haar measure by $\sigma_4(w,z)$.
The monomial integral is denoted by
\begin{equation}
\la\bm M\ra= \int\rmd\sigma_4(w,z)\; \cM_4(\bm w) \; .
\label{GS:defint}\end{equation}

\section{\label{O} Integration over \boldmath $O(d)$}


For any of the classical groups, the Haar measure is invariant under left-
and right-multiplication with a fixed group element $u$, and under taking the
inverse. For $O(d)$ these properties read:
\begin{equation}
\int\rmd\sigma_1(w)\; f(w) = \int\rmd\sigma_1(w)\; f(u\, w)
 = \int\rmd\sigma_1(w)\; f(w\, u) = \int\rmd\sigma_1(w)\; f(w^T) \; ,
\label{O:inv}\end{equation}
where $f(w)$ is an arbitrary integrable function of the matrix entries of $w$.
For a monomial integral $\la M\ra$, as defined in equation~(\ref{GO:defint}),
those operations translate into corresponding operations on the integer matrix
$M$. They can be used to bring the monomial integral into a more convenient
form. In other words, we use transposition and/or column permutations to 
collect all non-zero elements of $M$ in the first $R\le d$ columns, 
minimizing $R$.

The general idea is to write the integral in terms of the full 
$d^2$-dimensional Euclidean space of all matrix elements of $w$, and to 
implement the restriction to the group manifold by appropriately chosen 
$\delta$-functions. Since the monomial $\cM_1(w)$ contains matrix elements 
from only the first $R$ column vectors of $w$, the integration may be 
restricted to those, as will be shown with the help of the final result, below. 
Meanwhile, one may as well assume that $R=d$. Denoting the column vectors by 
$\vec w_1, \vec w_2, \ldots, \vec w_R$, one may write:
\begin{align}
\la M\ra &= \int\rmd\sigma_1(w) \; \prod_{i,\xi =1}^{d,R} w_{i\xi}^{M_{i\xi}} =
\frac{\cN_1^{(R)}(M)}{\cN_1^{(R)}(o)} \; , \label{O:avM}\\
\cN_1^{(R)}(M) &= \prod_{\xi=1}^R\left\{\int\rmd\Omega(\vec w_\xi) 
   \prod_{i=1}^d w_{i\xi}^{M_{i\xi}} \right\} \; 
\prod_{\mu <\nu}^R \delta(\la\vec w_\mu|\vec w_\nu\ra) \qquad
\int\rmd\Omega(\vec w_\xi) \propto \prod_{i=1}^d
   \left\{\int\rmd w_{i\xi}\right\} \delta(\|\vec w_\xi\|^2 -1) \; .
\label{O:cN}\end{align}
The matrix $\bm o$ in equation~(\ref{O:avM}) is a $d$-dimensional square matrix,
which contains only zeros. The subscript in the symbol $\cN_1^{(R)}$ indicates
the symmetry parameter $\beta$, which is equal to one in the orthogonal case. 
In equation~(\ref{O:cN}) the 
integration region is the product space of $R$ unit spheres with constant 
measure $\rmd\Omega(\vec w_\xi)$. The orthogonality 
and the normalization is implemented with the help of appropriately chosen 
$\delta$-functions.

To convince oneself that the right hand side of equation~(\ref{O:avM})
really yields the Haar measure, it is sufficient to realize that for $R=d$
the measure used in equation~(\ref{O:cN}) is indeed invariant under 
transposition as well as left- and right-multiplication with any other fixed
group element. Similar arguments also apply in the case of the unitary and 
the unitary symplectic group.

To evaluate the integral in equation~(\ref{O:cN}), we integrate over the 
last column vector $\vec w_R$. This integration can be done in closed form, 
and the result is a linear combination of monomials in the matrix elements of 
the first $R-1$ column vectors. This means, the resulting expression will be 
of the following form:
\begin{equation}
\cN_1^{(R)}(M) = \sum_K C(K)\; \cN_1^{(R-1)}(M+K) \qquad
\cN_1^{(R)}(o) = C(0)\; \cN_1^{(R-1)}(o) \; ,
\label{O:idea}\end{equation}
where $K$ is an integer matrix with non-zero elements in the first $R-1$ 
columns only. The sum $\sum_K$ runs over a finite number of such matrices,
and for $\cN_1^{(R-1)}(M+K)$ only the first $R-1$ columns (of $M$ and of $K$)
need to be taken into account. The second relation in equation~(\ref{O:idea}) 
takes care of the proper normalization of the measure. The ratio between both 
equations, yields the desired recurrence relation:
\begin{equation}
\la M^{(R)}\ra = \sum_K \tilde C(K)\; \la M^{(R-1)}+K\ra \; .
\end{equation}
We will eventually use the supscript of the form $M^{(R)}$ to indicate that 
only the first $R$ column vectors of the respective matrix are taken into 
account. An integral, such as the one in equation~(\ref{O:avM}) will be called
``$R$-vector integral''. With only minor changes, this terminology is 
also applied in the unitary and unitary symplectic case.

\paragraph{The one-vector integral}
The starting point for the recurrence relation is the one-vector integral.
It has first been computed by Ullah~\cite{Ullah64}. In our notation, it reads:
\begin{equation}
\la\vec m\ra = \int\rmd\sigma_1(\vec w)\; \prod_{i=1}^d w_i^{m_i}
 = \left(\ts\frac{d}{2}\right)_{\bar m/2}^{-1}\prod_{i=1}^d
      \left(\ts \frac{1}{2}\right)_{m_i/2} \; ,
\label{O:v1}\end{equation}
where $(z)_a= \Gamma(z+a)/\Gamma(z)$ is the Pochhammer symbol, and 
$\bar m= \sum_{i=1}^d m_i$. The integral $\la\vec m\ra$ is non-zero only if
all components $m_i$ are even numbers.

\paragraph{The \boldmath $R$-vector integral}
The recurrence relation for the $R$-vector case has been obtained 
in~\cite{Gor02}. It relates the desired $R$-vector integral to a linear 
combination of $($$R$$-$$1$$)$-vector integrals. To obtain an explicite result, 
this recurrence relation must be continued until $R=1$ is reached, where the 
one-vector result can be inserted.
\begin{align}
&\la M\ra = \chi_{\bar m:\text{even}}
   \sum_{\vec\kappa} {\vec m\choose\vec\kappa}\; \la\vec\kappa\ra\; 
   B\!\left(\ts \frac{\bar m}{2},\frac{\bar\kappa}{2};
      \frac{d}{2},\frac{R-1}{2}\right)\;
   \sum_K (\vec m-\vec\kappa\, | K)\; 
   \la \vec k_{\rm cs}\ra\; \la M^{(R-1)}+ K\ra  \; ,
\label{O:vR}\\
\text{where}\quad
&B(a,b;z_1,z_2) = (-)^{a-b}\; \frac{(z_1)_b\; (z_1)_{a-b}}{(z_1-z_2)_a} \; .
\label{O:defB}\end{align}
In this formula, $\vec m$ is the $R$-th column vector of $M$ and $\bar m$ is
the sum of all its entries. The prefactor $\chi_{\bar m:\text{even}}$ is one
if $\bar m$ is even, otherwise it is zero. The ``binomial'' of the integer 
vectors $\vec m$ and $\vec\kappa$ is just a short hand for the product of 
binomials of their respective components:
\begin{equation}
{\vec m\choose\vec\kappa} = \prod_{i=1}^d {m_i\choose \kappa_i} \; .
\label{O:shortbin}\end{equation}
The symbol, $\la\vec\kappa\ra$, denotes a one-vector
average, as defined in equation~(\ref{O:v1}). Due to those two quantities, the
summation is restricted to such $\vec\kappa$ for which
$\forall i\, :\, 0\le\kappa_i\le m_i\, ,\, \kappa_i :$even. Analogous to
$\bar m$, $\bar\kappa$ denotes the sum of all components of the integer vector 
$\vec\kappa$. The second sum in equation~(\ref{O:vR}) runs over the 
$d$-dimensional integer matrix $K$ with non-negative entries in the first $R-1$
columns, only. The following ``vector-multinomial'' is again a short hand,
which reads:
\begin{equation}
(\vec m-\vec\kappa\, | K) = \prod_{i=1}^d 
   (m_i-\kappa_i|K_{i1},\ldots,K_{i,R-1}) \; ,
\label{O:shortmul}\end{equation}
where the row sums in $K$ must be equal to $m_i - \kappa_i$. 
The vector $\vec k_{\rm cs}$ denotes the vector of column-sums of $K$. It has
at most $R-1$ non-zero components. Since our formula contains the one-vector
average of $\vec k_{\rm cs}$, the summation over $K$ should be restricted to
those $K$, which have only even column sums. The last term is a 
$($$R$$-$$1$$)$-vector average, where the matrix entries of $K$ have been
added to the matrix entries in the first $R-1$ columns of $M$.

As it turns out, the function $B(a,b;z_1,z_2)$ appears again
in the integration formulas for $U(d)$ and $\Sp(2d)$, although with slightly
different arguments.

\paragraph{Zero column vectors}
With the recurrence formula~(\ref{O:vR}) at hand, it is straight forward to
prove that the restriction of the integration in equation~(\ref{O:cN}) to the
first $R$ column vectors of $w$ is valid. This follows from the fact that the
recurrence relation yields $\la M\ra= \la M^{(R-1)}\ra$ if all components of 
the $R$'th column vector of $M$ are zero.

\paragraph{Vanishing integrals}
It is well known ({\it e.g.} reference~\cite{Bro81}), that the integral 
$\la M\ra$ vanishes if any column sum or row sum of $M$ gives an odd number. 
This can be directly seen from equation~(\ref{O:vR}), which requires $\bar m$ 
to be even in order to yield a non-zero result. Due to the invariance 
properties of the Haar measure, this statement applies to any column- or 
row-vector.

\section{\label{U} Integration over \boldmath $U(d)$}

For the unitary group, we denote the normalized Haar measure by $\sigma_2$.
It has analogous invariance properties as $\sigma_1$ of the orthogonal group, 
discussed in the previous section. In the present case, one finds for a fixed 
group element $u\in U(d)$:
\begin{equation}
\int\rmd\sigma_2(w)\; f(w,w^\dagger) = \int\rmd\sigma_2(w)\; 
  f(u\, w, w^\dagger\, u^\dagger) = \int\rmd\sigma_2(w)\; 
  f(w\, u,u^\dagger\, w^\dagger) = \int\rmd\sigma_2(w)\; f(w^\dagger,w) \; ,
\label{U:inv}\end{equation}
where $f(w,z)$ is an analytic integrable function of the matrix entries of
$w$ and $z$.
These operations translate into corresponding operations on the integer 
matrices $M,N$, which leave the monomial integral $\la N|M\ra$, 
equation~(\ref{GU:defint}), invariant. In particular: (i) simultanous 
column permutations: $\la N|M\ra = \la N\pi|M\pi\ra$, (ii) simultaneous row 
permutations: $\la N|M\ra = \la\pi N|\pi M\ra$, where $\pi$ is an arbitrary 
permutation matrix, and (iii) conjugate transposition:
$\la N|M\ra = \la M^T|N^T\ra$. Due to the invariance under column 
permutations, we may assume without loss of generality that the non-zero 
elements of $M$ and $N$ are all restricted to the first $R\le d$ 
columns. In other words, this means that the monomial $\cM_2(w)$ contains no
matrix elements from column vectors $\vec w_\mu$ with $\mu > R$.

Analogous to the case of the orthogonal group, we again write the integral in 
terms of the flat Euclidean space of all complex matrix elements of
$w\in U(d)$. Then, we implement the restriction to the group manifold by 
appropriately chosen $\delta$-functions: 
\begin{align}
\la N|M\ra &= \int\rmd\sigma_2(w) \; \prod_{i,\xi =1}^{d,R} 
   w_{i\xi}^{M_{i\xi}}\; (w^*_{i\xi})^{N_{i\xi}} =
\frac{\cN_2^{(R)}(M,N)}{\cN_2^{(R)}(o,o)} \; , \label{U:avM}\\
\cN_2^{(R)}(M,N) &= \prod_{\xi=1}^R\left\{\int\rmd\Omega_2(\vec w_\xi) 
   \prod_{i=1}^d w_{i\xi}^{M_{i\xi}}\; (w^*_{i\xi})^{N_{i\xi}} \right\} \; 
\prod_{\mu <\nu}^R \delta^{(2)}(\la\vec w_\mu|\vec w_\nu\ra) \; .
\label{U:cN}\end{align}
The subscript $2$ in the symbol $\cN_2^{(R)}$ stands for the unitary case 
$\beta=2$. The first $R$ column vectors of the unitary matrix $w$ are denoted
by $\vec w_1,\ldots\vec w_R$. The integration region in equation~(\ref{U:cN})
is the product space of $R$ unit spheres with constant measure 
$\rmd\Omega_2(\vec w_\xi)$:
\begin{equation}
\int\rmd\Omega_2(\vec w_\xi) \propto \prod_{i=1}^d
   \left\{\int\rmd^2(w_{i\xi})\right\} \delta(\|\vec w_\xi\|^2 -1) \; ,
\label{U:sphere}\end{equation}
where we define the flat measure on the complex plane via 
$z\in\mathbb{C}\, :\, \rmd^2(z)= \rmd({\rm Re}\, z)\, \rmd({\rm Im}\, z)$.
The $\delta$-functions in the equations~(\ref{U:cN}) and~(\ref{U:sphere}) 
implement the orthogonality conditions and the normalization. Note that the
$\delta$-function in equation~(\ref{U:cN}) really is the product of two
$\delta$-functions:
$\delta^{(2)}(z) = \delta({\rm Re}\, z)\; \delta({\rm Im}\, z)$.

Since we assumed that the monomial $\cM_2(w)$ contains matrix elements from 
only the first $R$ column vectors, we may restrict the integration 
in~(\ref{U:cN}) to those vectors. As in the orthogonal case, this does not
affect the result of the monomial integral, as will be shown below with the
help of the final result. Meanwhile, one may as well assume that $R=d$. Note 
that for $R=d$ the integration measure in equation~(\ref{U:cN}) is indeed 
invariant under the transformations performed in~(\ref{U:inv}). This guarantees 
that equation~(\ref{UR:res}) really yields the Haar measure.

\subsection{\label{U1}The one-vector formula}

In the one-vector case, the matrices $M$ and $N$ can be replaced by the
$d$-dimensional vectors $\vec m$ and $\vec n$. In that case, there are no 
orthogonality conditions to obey. We may write:
\begin{equation}
\la\vec n\, |\, \vec m\ra = \int\rmd\sigma_2(w)\; \prod_{i=1}^d w_i^{m_i}\; 
   (w^*_i)^{n_i}
 = \frac{\cN^{(1)}_2(\vec m, \vec n)}{\cN^{(1)}_2(\vec o,\vec o)} \qquad
\cN^{(1)}_2(\vec m,\vec n) = \int\rmd\Omega_2(\vec w)\; 
   \prod_{i=1}^d w_i^{m_i}\; (w^*_i)^{n_i} \; .
\label{U1:defint}\end{equation}
As suggested in equation~(\ref{U:sphere}), we integrate over the full space
$\mathbb{R}^{2d}$, while the normalization is implemented with the help of a 
$\delta$-function. This introduces an integration constant, denoted
by $C_2(d,1)$:
\begin{equation}
\cN^{(1)}_2(\vec m,\vec n) = C_2(d,1) \prod_{i=1}^d \left\{\int\rmd^2(w_i)\; 
  w_i^{m_i}\; (w^*_i)^{n_i}\right\} \delta\left(\|\vec w\|^2 - 1\right) \; .
\label{U1:eq2}\end{equation}
The $\delta$-function is removed as follows: Setting $w_i = u_i/\sqrt{r}$ we 
get:
\begin{equation}
\cN^{(1)}_2(\vec m,\vec n) \; r^{d+(\bar m+\bar n)/2-1} = C_2(d,1)
   \prod_{i=1}^d \left\{\int\rmd^2(u_i)\; u_i^{m_i}\; (u^*_i)^{n_i}\right\}
   \delta\left(\ts\sum_i |u_i|^2 - r\right) \; ,
\end{equation}
where $\bar m =\sum_{i=1}^d m_i$ and $\bar n =\sum_{i=1}^d n_i$. Multiplying
both sides of the equation with $\rme^{-r}$ and integrating on $r$ from $0$ to 
$\infty$, the $\delta$-function disappears:
\begin{equation}
\cN^{(1)}_2(\vec m,\vec n) \; \Gamma\left(d+\ts\frac{\bar m+\bar n}{2}\right)\; 
 = C_2(d,1) \prod_{i=1}^d \int\rmd^2(u_i)\; u_i^{m_i}\; (u^*_i)^{n_i}\;
   \rme^{- |u_i|^2} = \prod_{i=1}^d f(m_i,n_i,0) \; .
\label{U1:eq3}\end{equation}
The Gaussian integral $f(m,n,\alpha)$ is defined in 
equation~(\ref{A4:defint}) in the appendix. The general result for that
integral is given in equation~(\ref{A4:res}). For $\alpha=0$, it evaluates to
\begin{equation}
f(m,n,0)= \pi\; \delta_{m,n}\; n!\; .
\label{U1:fres}\end{equation}
From this, it follows:
\begin{equation}
\cN^{(1)}_2(\vec m,\vec n) \; \Gamma\left(d+\ts\frac{\bar m+\bar n}{2}\right)\;
 = C_2(d,1)\; \delta_{\vec m,\vec n} \prod_{i=1}^d 
      \left\{ \pi\; m_i! \right\} \qquad
\cN^{(1)}_2(\vec o,\vec o) \; \Gamma(d) = C_2(d,1)\; \pi^d \; ,
\end{equation}
where $\delta_{\vec m,\vec n}$ denotes the product of Kronecker deltas between
the components of $\vec m$ and $\vec n$. This finally leads to
\begin{equation}
\la\vec n\, |\, \vec m\ra =
\delta_{\vec m,\vec n}\; (d)^{-1}_{\bar m}\; \prod_{i=1}^d m_i! \; .
\label{U1:res}\end{equation}

\subsection{\label{UR}The $R$-vector formula}

To solve the $R$-vector integral, we start from equation~(\ref{U:cN}) and
separate the integration over the first $R-1$ column vectors from the 
integration over the last clumn vector $\vec w_R$. Let us denote the $R$-th
column vectors of $M,N$ and $w$ with $\vec m,\vec n$, and $\vec w$, 
respectively.
\begin{align}
\cN_2^{(R)}(M,N) &= \prod_{\xi=1}^{R-1}\left\{\int\rmd\Omega_2(\vec w_\xi)\; 
   \prod_{i=1}^d w_{i\xi}^{M_{i\xi}}\; (w^*_{i\xi})^{N_{i\xi}} \right\} 
   \prod_{\mu <\nu}^{R-1}\left\{ 
   \delta^{(2)}(\la\vec w_\mu|\vec w_\nu\ra )\right\}\; 
   {\cal J}_2^{(R)}(\vec m,\vec n) 
\label{UR:N}\\
{\cal J}_2^{(R)}(\vec m,\vec n) &= \int\rmd\Omega_2(\vec w) \prod_{i=1}^d
   \left\{ w_i^{m_i}\; (w^*_i)^{n_i} \right\} \prod_{\mu=1}^{R-1} 
   \delta^{(2)}(\la\vec w_\mu|\vec w\ra ) \; .
\label{UR:J}\end{align}
We start by flattening the integration measure $\Omega_2$, using the technique
from the one-vector integral. With 
$\rmd^2(w_i)= \rmd({\rm Re}\, w_i)\, \rmd({\rm Im}\, w_i)$ we may write:
\begin{equation}
{\cal J}_2^{(R)}(\vec m,\vec n) = C_2(d,R)\;
   \prod_{i=1}^d\left\{\int\rmd^2(w_i)\; w_i^{m_i}\; (w^*_i)^{n_i}
   \right\} \delta\left(\ts \sum_i w_i^* w_i -1\right) \prod_{\mu=1}^{R-1} 
   \delta^{(2)}\left(\la\vec w_\mu|\vec w\ra \right) \; .
\end{equation}
The transformation $w_i= u_i/\sqrt{r}$ leads to:
\begin{equation}
{\cal J}_2^{(R)}(\vec m,\vec n)\; r^{d+(\bar m+\bar n)/2-1-(R-1)} = C_2(d,R)\;
   \prod_{i=1}^d\left\{\int\rmd^2(u_i)\; u_i^{m_i}\; (u^*_i)^{n_i}\right\} 
   \delta\left(\ts \sum_i u_i^* u_i -r\right)
   \prod_{\mu=1}^{R-1} \delta^{(2)}\left(\la\vec w_\mu| \vec u\ra \right) \; .
\end{equation}
Mulitplying both sides with $\rme^{-r}$ and integrating on $r$ from $0$ to 
$\infty$ gives:
\begin{equation}
{\cal J}_2^{(R)}(\vec m,\vec n)\; 
   \Gamma\left(\ts d-R+1 +\frac{\bar m+\bar n}{2}\right) = C_2(d,R)\; 
   \prod_{i=1}^d\left\{\int\rmd^2(u_i)\; u_i^{m_i}\; (u^*_i)^{n_i}\; 
   \rme^{-u_i^* u_i} \right\} \prod_{\mu=1}^{R-1}
   \delta^{(2)}\left(\la\vec w_\mu| \vec u\ra \right) \; .
\end{equation}
Using the Fourier representation of the one-dimensional delta function, we 
may write:
\begin{equation}
\delta^{(2)}(w) = \iint\rmd x\,\rmd y\; 
   \rme^{2\pi\rmi\, (x\, {\rm Im}\,w + y\, {\rm Re}w)} 
 = \iint\frac{\rmd x\,\rmd y}{\pi^2}\; \rme^{x (w-w^*) + \rmi y (w+w^*)}
 = \int\frac{\rmd^2(z)}{\pi^2}\; \rme^{wz-w^* z^*}
 = \int\frac{\rmd^2(z)}{\pi^2}\; \rme^{2\rmi\, {\rm Im}(wz)} \; .
\label{UR:del2}\end{equation}
With the help of this representation, we write:
\begin{equation}
\prod_{\mu=1}^{R-1}\delta^{(2)}\left(\ts \sum_i w_{i\mu}^*\, u_i\right) = 
   \prod_{\mu=1}^{R-1} \int\frac{\rmd^2(\tau_\mu)}{\pi^2}\; \exp\left(\ts 
   \tau_\mu\, \sum_i w_{i\mu}^* u_i - \tau_\mu^*\, \sum_i w_{i\mu} u_i^*
   \right)\; .
\end{equation}
Therefore
\begin{align}
{\cal J}_2^{(R)}(\vec m,\vec n) &= 
   \frac{C_2(d,R)}{\Gamma\left(d-R+1 +\frac{\bar m+\bar n}{2}\right)}
   \prod_{\mu=1}^{R-1}\left\{\int\frac{\rmd^2(\tau_\mu)}{\pi^2}\right\}\; 
   \prod_{i=1}^d f(m_i,n_i,\alpha_i)\qquad
\alpha_i = \sum_{\mu=1}^{R-1} \tau_\mu\, w_{i\mu}^*
\label{UR:J2}\\ 
f(m,n,\alpha) &= \int\rmd^2(u)\; u^m\; (u^*)^n\; \rme^{-u^* u}\; 
   \rme^{\alpha\, u - \alpha^* u^*} \notag\\
&= \pi\; (-)^m \sum_{\kappa=0}^p (-)^\kappa\; \kappa!\; {m \choose \kappa}\; 
   {n \choose \kappa}\; (\alpha^*)^{m-\kappa}\; \alpha^{n-\kappa}\; 
   \rme^{-\alpha^*\alpha} \qquad 
p= \min(m,n)\; .
\notag\end{align}
For details about the computation of that integral, see appendix. In 
order to put the product of the functions $f(m_i,n_i,\alpha_i)$ into a 
suitable form, we define additional integer vectors $\vec\kappa$ and 
$\vec p$, where $p_i = \min(m_i,n_i)$:
\begin{equation}
\prod_{i=1}^d f(m_i,n_i,\alpha_i) = \pi^d\; (-)^{\bar m}\; 
   \sum_{\vec\kappa = \vec o}^{\vec p} (-)^{\bar\kappa}
   \prod_{i=1}^d\left\{ {m_i \choose \kappa_i}\; {n_i \choose \kappa_i}\; 
   \kappa_i!\; (\alpha_i^*)^{m_i-\kappa_i}\; \alpha_i^{n_i-\kappa_i} 
   \right\}\; \rme^{-\sum_i |\alpha_i|^2} \; .
\end{equation}
With $\bar m$ and $\bar\kappa$, we denote the sum of vector components in 
$\vec m$ and $\vec\kappa$, respectively. For the expansion of the powers of the 
coefficients $\alpha_i$ and $\alpha_i^*$, we need the integer matrices $K$ and 
$L$. Both are $d$-dimensional matrices with non-zero elements in the first 
$R-1$ columns, only.
\begin{align}
\prod_{i=1}^d (\alpha_i^*)^{m_i-\kappa_i} &= \prod_{i=1}^d 
   \left(\ts \sum_{\mu=1}^{R-1} \tau_\mu^*\, w_{i\mu}\right)^{m_i-\kappa_i}
 = \sum_K (\vec m-\vec\kappa|K)\; \prod_{\mu=1}^{R-1}\prod_{i=1}^d 
   (\tau_\mu^*)^{K_{i\mu}}\, w_{i\mu}^{K_{i\mu}} \notag\\
&= \sum_K (\vec m-\vec\kappa|K)\; \prod_{\mu=1}^{R-1}
   (\tau_\mu^*)^{\bar k_\mu}\; \prod_{i=1}^d w_{i\mu}^{K_{i\mu}} \; ,
\end{align}
where $\vec k_1,\ldots\vec k_{R-1}$ denote the column vectors of $K$, while
the overbars denote, as usual, the summation over all vector components. In a
completely analogous way, we write also:
\begin{equation}
\prod_{i=1}^d \alpha_i^{n_i-\kappa_i} = \sum_L (\vec n-\vec\kappa|L)\; 
   \prod_{\mu=1}^{R-1} \tau_\mu^{\bar l_\mu}\; \prod_{i=1}^d 
   (w_{i\mu}^*)^{L_{i\mu}} \; ,
\end{equation}
where $\vec l_1,\ldots\vec l_{R-1}$ denote the column vectors of $L$.
Thus, we obtain from equation~(\ref{UR:J2}):
\begin{align}
{\cal J}_2^{(R)}(\vec m,\vec n) &= \frac{C_2(d,R)\; \pi^d\; (-)^{\bar m}}
   {\Gamma\left(d-R+1 +\frac{\bar m+\bar n}{2}\right)}\; 
   \sum_{\vec\kappa = \vec o}^{\vec p} (-)^{\bar\kappa}\; \prod_{i=1}^d\left\{
   {m_i \choose \kappa_i}\; {n_i \choose \kappa_i}\; \kappa_i! \right\} 
\notag\\
&\qquad\times \sum_{K,L} (\vec m-\vec\kappa|K)\; (\vec n-\vec\kappa|L)\; 
   \prod_{\mu=1}^{R-1}\int\frac{\rmd^2(\tau_\mu)}{\pi^2}\;
   (\tau_\mu^*)^{\bar k_\mu}\; \tau_\mu^{\bar l_\mu}\; 
   \rme^{-\sum_i |\alpha_i|^2} \;
      \prod_{i=1}^d w_{i\mu}^{K_{i\mu}} \; (w_{i\mu}^*)^{L_{i\mu}} \; .
\end{align}
Due to the fact that the integration in equation~(\ref{U:cN}) is restricted
to column vectors $\vec w_\xi$ which are pairwise orthogonal, we obtain:
\begin{equation}
\sum_i |\alpha_i|^2 = \sum_\mu\sum_\nu \tau_\mu\, \tau_\nu^*\;
   \sum_i w_{i\nu}\, w_{i\mu}^* = \sum_\mu |\tau_\mu|^2 \; .
\end{equation}
This is a crucial point in our calculation. It eliminates the remaining
matrix elements of $w$ from the exponential, and 
${\cal J}_2^{(R)}(\vec m,\vec n)$ turns into a linear combination of 
simple monomials. The integrals over $\tau_\mu$ are now easily evaluated, using 
equation~(\ref{U1:eq3}) from the calculation of the one-vector average:
\begin{align}
{\cal J}_2^{(R)}(\vec m,\vec n) &= \frac{C_2(d,R)\; \pi^{d+R-1}\; (-)^{\bar m}}
   {\Gamma\left(d-R+1 +\frac{\bar m+\bar n}{2}\right)}\;
   \sum_{\vec\kappa = \vec o}^{\vec p} (-)^{\bar\kappa}\prod_{i=1}^d\left\{ 
   {m_i \choose \kappa_i}\; {n_i \choose \kappa_i}\; \kappa_i! \right\} 
\notag\\
&\qquad\times \sum_{K,L} (\vec m-\vec\kappa|K)\; (\vec n-\vec\kappa|L)\;
   \prod_{\mu=1}^{R-1} \delta_{\bar k_\mu,\bar l_\mu}\; \bar k_\mu!\;
   \prod_{i=1}^d w_{i\mu}^{K_{i\mu}} \; (w_{i\mu}^*)^{L_{i\mu}} \; .
\end{align}
Now, we insert this result into equation~(\ref{UR:N}) and write for 
$\cN_2^{(R)}(M,N)$ and $\cN_2^{(R)}(o,o)$:
\begin{align}
\cN_2^{(R)}(M,N) &= \frac{C_2(d,R)\; \pi^{d-R+1}\; (-)^{\bar m}}
   {\Gamma\left(d-R+1 +\frac{\bar m+\bar n}{2}\right)}\;
   \sum_{\vec\kappa = \vec o}^{\vec p} (-)^{\bar\kappa}\; \prod_{i=1}^d\left\{
   {m_i \choose \kappa_i}\; {n_i \choose \kappa_i}\; \kappa_i! \right\} 
 \sum_{K,L} (\vec m-\vec\kappa|K)\; (\vec n-\vec\kappa|L)\notag\\
&\qquad\times \prod_{\mu=1}^{R-1}\left\{ \delta_{\bar k_\mu,\bar l_\mu}\; 
   \bar k_\mu!\right\}\quad
   \prod_{\xi=1}^{R-1}\left\{\int\rmd\Omega_2(\vec w_\xi)\; \prod_{i=1}^d 
   w_{i\xi}^{M_{i\xi}+K_{i\xi}}\; (w^*_{i\xi})^{N_{i\xi}+L_{i\xi}}
   \right\} \left\{\prod_{\mu <\nu}^{R-1} \delta(\la\vec w_\mu|\vec w_\nu\ra)
   \right\} \notag\\
&= \frac{C_2(d,R)\; \pi^{d-R+1}\; (-)^{\bar m}}
   {\Gamma\left(d-R+1 +\frac{\bar m+\bar n}{2}\right)}\;
   \sum_{\vec\kappa = \vec o}^{\vec p} (-)^{\bar\kappa}\; \prod_{i=1}^d\left\{
   {m_i \choose \kappa_i}\; {n_i \choose \kappa_i}\; \kappa_i!\right\} \notag\\
&\qquad\times
   \sum_{K,L} (\vec m-\vec\kappa|K)\; (\vec n-\vec\kappa|L)\;
   \prod_{\mu=1}^{R-1}\left\{\delta_{\bar k_\mu,\bar l_\mu}\; 
   \bar k_\mu!\right\}\; \cN_2^{(R-1)}(M^{(R-1)}+K,N^{(R-1)}+L) \\
\cN_2^{(R)}(o,o) &= \frac{C_2(d,R)\, \pi^{d-R+1}}
   {\Gamma(d-R+1)}\; \cN_2^{(R-1)}(o,o) \; .
\end{align}
Thus we obtain:
\begin{align}
\la N|M\ra &= (-)^{\bar m}\; 
   (d-R+1)^{-1}_{(\bar m+\bar n)/2}\;
   \sum_{\vec\kappa = \vec o}^{\vec p} (-)^{\bar\kappa}\; \prod_{i=1}^d\left\{
   {m_i \choose \kappa_i}\; {n_i \choose \kappa_i}\; \kappa_i!\right\} \notag\\
&\qquad\times
   \sum_{K,L} (\vec m-\vec\kappa|K)\; (\vec n-\vec\kappa|L)\;
   \prod_{\mu=1}^{R-1}\left\{\delta_{\bar k_\mu,\bar l_\mu}\;
   \bar k_\mu! \right\}\; \la N^{(R-1)}+L|M^{(R-1)}+K\ra \; .
\end{align}
Note that due to the multinomial coefficients $(\vec m-\vec\kappa|K)$ and 
$(\vec n-\vec\kappa|L)$, for the column sums of $K$ and $L$, it holds that:
$\sum_{\mu=1}^{R-1} \bar k_\mu = \bar m - \bar\kappa$, and also 
$\sum_{\mu=1}^{R-1} \bar l_\mu = \bar n - \bar\kappa$. In addition, to obtain
a non-zero contribution, the column sums of $K$ and $L$ must be the same also:
$\forall 1\le\mu\le R-1 \, :\, \bar k_\mu = \bar l_\mu$. Together, both 
conditions imply that for $\bar m\ne \bar n$, the whole monomial integral 
gives zero. Therefore:
\begin{align}
\la N|M\ra &= \delta_{\bar m,\bar n}\; \sum_{\vec\kappa = \vec o}^{\vec p}
   {\vec m\choose \vec\kappa}\; {\vec n\choose \vec\kappa}\; 
   \la\vec\kappa|\vec\kappa\ra\; B(\bar m,\bar\kappa; d,R-1)\notag\\
&\qquad\times \sum_{K,L} (\vec m-\vec\kappa|K)\; (\vec n-\vec\kappa|L)\; 
   \la\vec l_{\rm cs}|\vec k_{\rm cs}\ra\; \la N^{(R-1)}+L|M^{(R-1)}+K\ra \; .
\label{UR:res}\end{align}
In this formula, we use the vector notation for the product of binomials and
multinomials, as introduced in section~\ref{O}. We also express the products 
$\prod_{i=1}^d\, \kappa_i!$ and 
$\prod_{\mu=1}^{R-1}\, \delta_{\bar k_\mu,\bar l_\mu}\, \bar k_\mu!$ in terms
of the one-vector averages, given in equation~(\ref{U1:res}). In this way, the 
coefficient $B(\bar m,\bar\kappa; d,R-1)$, defined in equation~(\ref{O:defB}), 
comes into play. The summation over $\vec\kappa$ is restricted to those
vectors, for which $\forall i\, :\, 0\le \kappa_i\le \min(m_i,n_i)$. $K$ and
$L$ are $d$-dimensional matrices with non-negative integer entries, which have
non-zero elements in their first $R-1$ columns, only. The summation over $K$ 
is restricted to those $K$, for which 
$\forall i\, :\, \sum_\mu K_{i\mu} = m_i-\kappa_i$. Similarly, the summation
over $L$ is restricted to those $L$, for which
$\forall i\, :\, \sum_\mu L_{i\mu} = n_i -\kappa_i$. In addition, the 
one-vector average $\la\vec l_{\rm cs}|\vec k_{\rm cs}\ra$ implies that the
column-sums of $K$ must agree with the corresponding column sums of $L$. These
column sums are arranged in the first $R-1$ components of the $d$-dimensional
vectors $\vec k_{\rm cs}$ and $\vec l_{\rm cs}$, respectively.

\paragraph{Zero column vectors}
Assume that in the monomial integral $\la N|M\ra$, the $R$'th column of $M$ 
and $N$ contain both only zeros. Equation~(\ref{UR:res}) then yields:
$\la N|M\ra = \la N^{(R-1)}|M^{(R-1)}\ra$. This proves by a simple induction
argument, that the integration space in ~(\ref{U:cN}) may indeed be restricted
to those column vectors, which contain matrix elements from the monomial
$\cM_2(w)$.

\paragraph{Vanishing integrals}
Denoting monomials in the form:
$w_{a_1 b_1}^*\, \ldots\, w_{a_p b_p}^*\, w_{\alpha_1 \beta_1}\, \ldots\,
  w_{\alpha_q \beta_q}$ (see for instance, reference~\cite{BroBee96}), it 
has been shown that the monomial integral 
$\la w_{a_1 b_1}^*\, \ldots\, w_{a_p b_p}^*\, w_{\alpha_1 \beta_1}\, \ldots\, 
  w_{\alpha_q \beta_q}\ra$ vanishes, unless $p=q$ and 
$\alpha_1,\ldots,\alpha_q$ is a permutation of $a_1,\ldots,a_p$ and 
$\beta_1,\ldots,\beta_q$ is a permutation of $b_1,\ldots, b_p$. This is 
equivalent to the requirement that all columns sums and row sums of $M$ 
agree with the corresponding column sums and row sums of $N$. In view of the 
invariance of the Haar measure under row- or column permutations, this is a 
trivial consequence of the leading factor $\delta_{\bar m,\bar n}$ in 
equation~(\ref{UR:res}).

\section{\label{S} Integration over \boldmath $\Sp(2d)$}

For the unitary symplectic group, we denote the normalized Haar measure by
$\sigma_4$. For
\begin{equation}
{\it Sp}(2d) \ni \bm w= \begin{pmatrix} z^* & w\\ -w^* & z\end{pmatrix}\; ,
\quad\text{let}\;\; f(\bm w)= f(w,z,w^*,z^*)\; ,
\end{equation}
where $f(w_1,z_1,w_2,z_2)$ is an arbitrary analytic(?) and integrable
function of the matrix elements of $w_1$, $z_1$, $w_2$, and $z_2$. The 
invariance properties of the Haar measure then imply:
\begin{equation}
\int\rmd\sigma_4(w,z)\; f(\bm w) = \int\rmd\sigma_4(w,z)\; f(\bm w\, \bm u)
 = \int\rmd\sigma_4(w,z)\; f(\bm u\, \bm w) = \int\rmd\sigma_4(w,z)\; 
   f(\bm w^\dagger) \; ,
\label{S:inv}\end{equation}
where $\bm u$ is a arbitary but fixed element of $\Sp(2d)$.

Since we are interested in the case, where the function $f$ is a monomial, we 
use some particular permutation matrices from $\Sp(2d)$ to obtain invariance 
relations for the monomial integral $\la\bm M\ra$, under certain 
transformations of $\bm M$. For the following considerations, let
\begin{equation}
\bm w= \begin{pmatrix} z^* & w\\-w^* & z\end{pmatrix}\; , \qquad
\bm u= \begin{pmatrix} v^* & u\\-u^* & v\end{pmatrix} \; .
\end{equation}
\begin{itemize}
\item[(1)] For $u= \pi$, a $d$-dimensional permutation matrix, and $v= o$,
multiplication from the left leads to
\begin{equation}
\bm w \quad\to\quad \bm w\, \bm u=
\begin{pmatrix} -w\, \pi & z^*\,\pi\\ -z\, \pi & -w^*\,\pi\end{pmatrix} \; ,
\end{equation}
such that 
\begin{align}
&\la\bm M\ra= \lla\begin{pmatrix} N' & M\\ N & M'\end{pmatrix}\rra
 = \int\rmd\sigma_4(\bm w)\; \prod_{i,j=1}^d w_{ij}^{M_{ij}} z_{ij}^{M'_{ij}}
      (w_{ij}^*)^{N_{ij}} (z_{ij}^*)^{N'_{ij}} \notag\\
&\quad\to\quad \int\rmd\sigma_4(\bm w)\; \prod_{i,j=1}^d (z^*\pi)_{ij}^{M_{ij}}
   (-w^*\pi)_{ij}^{M'_{ij}} (z\pi)_{ij}^{N_{ij}} (-w\pi)_{ij}^{N'_{ij}}
 = (-)^{\bar M' +\bar N'} \lla\begin{pmatrix} M\, \pi^{-1} & N'\, \pi^{-1}\\
      M'\,\pi^{-1} & N\, \pi^{-1}\end{pmatrix}\rra \; ,
\label{S:inv1}\end{align}
where $\bar M'$ and $\bar N'$ are the sum of all matrix entries in $M$ and $N$,
respectively. Similarly, multiplication with $\bm u$ from the right yields:
\begin{equation}
\la\bm M\ra \quad\to\quad \int\rmd\sigma_4\; \prod_{i,j=1}^d 
   (\pi z)_{ij}^{M_{ij}} (-\pi w)_{ij}^{M'_{ij}} (\pi z^*)_{ij}^{N_{ij}}
   (-\pi w^*)_{ij}^{N'_{ij}} = (-)^{\bar M' +\bar N'}\;
   \lla\begin{pmatrix} \pi^{-1} N & \pi^{-1} M'\\ \pi^{-1}N' & \pi^{-1}M
      \end{pmatrix}\rra \; .
\label{S:inv2}\end{equation}

\item[(2)] For $u= o$ and $v= \pi$, multiplication from the left yields:
\begin{equation}
\la\bm M\ra \quad\to\quad \int\rmd\sigma_4\; \prod_{i,j=1}^d
   (w \pi)_{ij}^{M_{ij}} (z \pi)_{ij}^{M'_{ij}} (w^* \pi)_{ij}^{N_{ij}}
   (z^* \pi)_{ij}^{N'_{ij}} = \lla\begin{pmatrix} N' \pi^{-1} & M \pi^{-1}\\ 
      N \pi^{-1} & M' \pi^{-1}\end{pmatrix}\rra \; ,
\label{S:inv3}\end{equation}
whereas multiplication from the right yields:
\begin{equation}
\la\bm M\ra \quad\to\quad \lla\begin{pmatrix}\pi^{-1} N' & \pi^{-1} M\\ 
      \pi^{-1} N & \pi^{-1} M'\end{pmatrix}\rra \;  .
\label{S:inv4}\end{equation}

\item[(3)] Finally, replacing $\bm w$ with its inverse, $\bm w^\dagger$,
yields:
\begin{equation}
\la\bm M\ra \quad\to\quad\int\rmd\sigma_4(\bm w)\; \prod_{i,j=1}^d
   (-w_{ji})^{M_{ij}} (z^*_{ji})^{M'_{ij}} (-w_{ji}^*)^{N_{ij}} 
   (z_{ji})^{N'_{ij}} = (-)^{\bar M +\bar N}\;
   \lla\begin{pmatrix} M^{\prime T} & M^T\\ N^T & N^{\prime T}\end{pmatrix}
\rra \; ,
\label{S:inv5}\end{equation}
where $\bar M$ and $\bar N$ are the sums of all matrix entries of $M$ and $N$,
respectively.
\end{itemize}
As we will show below [see equation~(\ref{S:vcond})], the monomial integral 
$\la\bm M\ra$ is zero, unless $\bar M +\bar M' = \bar N +\bar N'$. 
Equation~(\ref{S:inv5}) also implies that $\la\bm M\ra$ is zero, unless 
$\bar M +\bar N' = \bar N + \bar M'$. Both conditions together imply that 
\begin{equation}
\la\bm M\ra \ne 0 \quad\Rightarrow\quad \bar M =\bar M' =\bar N =\bar N' \; .
\end{equation}
We can therefore savely ignore the sign-prefactors in (\ref{S:inv1}),
(\ref{S:inv2}), and (\ref{S:inv5}). The results in item~(2) prove the 
invariance of $\la\bm M\ra$ under synchronous column permutations 
[equation~(\ref{S:inv3})] and synchronous row permutations 
[equation~(\ref{S:inv4})], applied to all submatrices of $\bm M$. 
Therefore, the equations~(\ref{S:inv1}) and~(\ref{S:inv2}) add only two 
additional operations: 
\begin{equation}
\begin{pmatrix} N' & M\\ N & M'\end{pmatrix} \quad\rightarrow\quad
\begin{pmatrix} M & N'\\ M' & N\end{pmatrix} \; , \qquad
\begin{pmatrix} N' & M\\ N & M'\end{pmatrix} \quad\rightarrow\quad
\begin{pmatrix} N & M'\\ N' & M\end{pmatrix} \; .
\end{equation}

\subsubsection*{The integration}

It is again possible to write the integral over ${\it Sp}(2d)$
in terms of an integral over the flat Euclidean space of all complex matrix 
elements of $w$ and $z$, implementing the restriction to the group 
manifold by appropriately chosen $\delta$-functions: 
\begin{equation}
\la\bm M\ra = \int\rmd\sigma_4(w,z)\; \cM_4(\bm w) =
   \frac{\cN_4^{(R)}(\bm M)}{\cN_4^{(R)}(\bm o)}\; , \qquad
\cM_4(\bm w) = \prod_{i,\xi =1}^{d,R} 
   w_{i\xi}^{M_{i\xi}}\; z_{i\xi}^{M'_{i\xi}}\; (w^*_{i\xi})^{N_{i\xi}}\;
   (z^*_{i\xi})^{N'_{i\xi}} \; .
\label{S:avM}\end{equation}
Here, we assume that the monomial $\cM_4(\bm w)$ contains no matrix elements
from the column vectors $\vec w_\mu, \vec z_\mu$ with $\mu > R$. One may have
used the invariance of $\la\bm M\ra$ under the above discussed transformations
in order to minimize $R$. The matrix $\bm o$ is of the same shape as $\bm M$,
but all its entries are equal to zero.
\begin{equation}
\cN_4^{(R)}(\bm M) = \left\{\prod_{\xi=1}^R\int\rmd
   \Omega_4(\vec w_\xi,\vec z_\xi)\right\}\; \cM_4(\bm w)\;
   \prod_{\mu <\nu}^R 
   \delta^{(2)}(\la\vec w_\mu|\vec w_\nu\ra + \la\vec z_\mu|\vec z_\nu\ra)\; 
\delta^{(2)}(\la\vec z^*_\mu|\vec w_\nu\ra - \la\vec w^*_\mu|\vec z_\nu\ra)\; .
\label{S:cN}\end{equation}
The subscript in the symbol $\cN_4^{(R)}$ is the symmetry parameter 
$\beta= 4$, since we consider here the unitary symplectic case. The first $R$ 
column vectors of the matrices $w,z$ are denoted by 
$\vec w_1,\ldots\vec w_R$, and $\vec z_1,\ldots\vec z_R$, respectively. The 
integration region in equation~(\ref{S:cN}) is the product space of $R$ unit 
hyperspheres with constant measure $\rmd\Omega_4(\vec w_\xi,\vec z_\xi)$:
\begin{equation}
\int\rmd\Omega_4(\vec w_\xi,\vec z_\xi) \propto \prod_{i=1}^d\left\{\int
   \rmd^4(w_{i\xi},z_{i\xi})\right\} 
   \delta(\|\vec w_\xi\|^2 + \|\vec z_\xi\|^2-1) \qquad
w,z\in\mathbb{C}\, :\, \rmd^4(w,z)= \rmd^2(w)\;\rmd^2(z) \; ,
\label{S:sphere}\end{equation}
where $\rmd^2(z)$ denotes the flat measure on the complex plane, as already
used in equation~(\ref{U:sphere}). The $\delta$-functions in~(\ref{S:cN}) 
and~(\ref{S:sphere}) implement the orthogonality conditions and the 
normalization, as defined in equation~(\ref{GS:ONrels}).

According to the assumption, the monomial $\cM_4(\bm w)$ contains matrix 
elements from the first $R$ columns of $w$ and $z$, only. Therefore, we 
restrict the integration in equation~(\ref{S:cN}) to those column vectors. The 
fact that this does not affect the result of the integral, will be shown only 
at the end of this section, where we have the final result, 
equation~(\ref{SR:res}) at our disposal. Note that the integration measure 
in~(\ref{S:cN}) is indeed invariant under the transformations performed 
in~(\ref{S:inv}). This guarantees that the above construction really implements 
the Haar measure.

\subsection{\label{S1} The one-vector formula}

In the one-vector case, the monomial $\cM_4(\bm w)$ contains matrix elements
from one column of $\bm w$ only. This means that the matrices $M,M',N',N$
have only zero entries in all columns, except for one which must be the same
for all matrices. We denote the column vectors with non-zero entries by
$\vec m, \vec m', \vec n$, and $\vec n'$ and use the following notation for 
the monomial integral:
\begin{equation}
\la\bm M\ra= \lla\left.\begin{matrix} \vec n'\\ \vec n\end{matrix}\right|
   \begin{matrix} \vec m\\ \vec m'\end{matrix}\rra 
= \int\rmd\Omega_4(\vec w,\vec z)\; \prod_{i=1}^d w_i^{m_i} z_i^{m'_i} 
   (w_i^*)^{n_i} (z_i^*)^{n'_i} \; .
\end{equation}
In order to evaluate that integral, we observe that there are no orthogonality 
relations to obey. The integral must therefore be equal to the one-vector 
integral $\la\vec q\, |\, \vec p\ra$ over the unitary group $U(2d)$, where 
$\vec p$ and $\vec q$ are the concatenations of the vectors $\vec m, \vec m'$ 
and $\vec n, \vec n'$, respectively. Therefore:
\begin{equation}
\lla\left.\begin{matrix} \vec n'\\ \vec n\end{matrix}\right|
   \begin{matrix} \vec m\\ \vec m'\end{matrix}\rra 
 = \delta_{\vec m,\vec n}\; \delta_{\vec m',\vec n'}\; 
      (2d)_{\bar m +\bar m'}^{-1} \prod_{i=1}^d m_i!\; m'_i! \; ,
\label{S1:res}\end{equation}
where $\bar m$ and $\bar m'$ denote the sum of all components of $\vec m$ and
$\vec m'$, respectively.

\subsection{\label{SR} The $R$-vector formula}

We aim at a recursion formula, which relates an $R$-vector integral to a 
linear combination of simpler $($$R$$-$$1$$)$-vector integrals. To derive 
such a relation, we separate the integration over the $R$-th column vectors 
of $w$ and $z$ from the rest. Let us denote the $R$-th column vectors
of $M,M',N,N'$ by $\vec m,\vec m',\vec n,\vec n'$. Then we may write:
\begin{align}
&\cN_4^{(R)}(\bm M) = \prod_{\xi=1}^{R-1}\left\{
   \int\rmd\Omega_4(\vec w_\xi, \vec z_\xi)\;
   \prod_{i=1}^d w_{i\xi}^{M_{i\xi}}\; z_{i\xi}^{M'_{i\xi}}
   (w^*_{i\xi})^{N_{i\xi}} (z^*_{i\xi})^{N'_{i\xi}}\;  \right\} \notag\\
&\qquad\qquad\qquad\times\; \prod_{\mu <\nu}^{R-1}\left\{ 
   \delta^{(2)}(\la\vec w_\mu|\vec w_\nu\ra + \la\vec z_\mu|\vec z_\nu\ra) \;
 \delta^{(2)}(\la\vec z^*_\mu|\vec w_\nu\ra - \la\vec w^*_\mu|\vec z_\nu\ra )
 \right\}\; {\cal J}_4^{(R)}(\vec m,\vec m',\vec n,\vec n') 
\label{SR:N}\\
&{\cal J}_4^{(R)}(\vec m,\vec m',\vec n,\vec n') = 
   \int\rmd\Omega_4(\vec w,\vec z)\; \prod_{i=1}^d\left\{ 
   w_i^{m_i} z_i^{m'_i} (w^*_i)^{n_i} (z_i^*)^{n'_i} \right\}\;
   \prod_{\mu=1}^{R-1} 
   \delta^{(2)}(\la\vec w_\mu|\vec w\ra + \la\vec z_\mu|\vec z\ra ) \;
   \delta^{(2)}(\la\vec z^*_\mu|\vec w\ra - \la\vec w^*_\mu|\vec z\ra )\; .
\label{SR:J}\end{align}
We start by flattening the integration measure $\Omega_4$. With 
equation~(\ref{S:sphere}) and
$\rmd^2(z)= \rmd({\rm Re}\,z)\, \rmd({\rm Im}\, z)$, we may write:
\begin{align}
{\cal J}_4^{(R)}(\ldots) &= C_4(d,R)\; \prod_{i=1}^d
   \left\{\int\rmd^2(w_i)\int\rmd^2(z_i)\; w_i^{m_i} z_i^{m'_i}
   (z_i^*)^{n'_i} (w^*_i)^{n_i} \right\}\; 
   \delta\left(\ts \sum_i |w_i|^2+ \sum_i |z_i|^2 -1\right)\notag\\
&\qquad\qquad \prod_{\mu=1}^{R-1}
   \delta^{(2)}(\la\vec w_\mu|\vec w\ra + \la\vec z_\mu|\vec z\ra ) \; 
   \delta^{(2)}(\la\vec z^*_\mu|\vec w\ra - \la\vec w^*_\mu|\vec z\ra )\; .
\end{align}
The delta function for the normalisation can be removed, using similar steps
as in the case of the unitary group. This yields:
\begin{align}
&{\cal J}_4^{(R)}(\ldots)= \frac{C_4(d,R)}
  {\Gamma\left(2(d-R+1) + (\bar m +\bar m' +\bar n' +\bar n)/2\right)}\;
   \prod_{i=1}^d\left\{\int\rmd^2(u_i)\int\rmd^2(v_i)\; 
   u_i^{m_i} v_i^{m'_i} (v^*_i)^{n'_i} (u^*_i)^{n_i}\;
   \rme^{- |u_i|^2 - |v_i|^2} \right\} \notag\\
&\qquad\times \prod_{\mu=1}^{R-1} 
   \delta^{(2)}(\la\vec w_\mu|\vec u\ra + \la\vec z_\mu|\vec v\ra ) \; 
   \delta^{(2)}(\la\vec z^*_\mu|\vec u\ra - \la\vec w^*_\mu|\vec v\ra )\; .
\label{SR:eq1}\end{align}
According to equation~(\ref{UR:del2}), the Fourier representations for the
remaining delta functions may be written as
\begin{align}
&\prod_{\mu=1}^{R-1}
   \delta^{(2)}\left[\ts \sum_i (w_{i\mu}^* u_i+ z_{i\mu}^* v_i)\right] = 
   \prod_{\mu=1}^{R-1}\int\frac{\rmd^2(\sigma_\mu)}{\pi^2}\; \prod_{i=1}^d 
   \rme^{2\rmi\, {\rm Im}[\sigma_\mu (w_{i\mu}^* u_i + z_{i\mu}^* v_i)]} 
= \prod_{\mu=1}^{R-1}\left\{ \int\frac{\rmd^2(\sigma_\mu)}{\pi^2} \right\}
  \prod_{i=1}^d \rme^{2\rmi\, {\rm Im}(\alpha_i u_i + \beta_i v_i)}
\notag\\
&\prod_{\mu=1}^{R-1}
   \delta^{(2)}\left[\ts \sum_i (z_{i\mu} u_i - w_{i\mu} v_i)\right] =
   \prod_{\mu=1}^{R-1}\int\frac{\rmd^2(\tau_\mu)}{\pi^2}\;
   \prod_{j=1}^d \rme^{2\rmi\, {\rm Im}[ \tau_\mu (z_{i\mu} u_i
      - w_{i\mu} v_i)]}
= \prod_{\mu=1}^{R-1}\left\{\int\frac{\rmd^2(\tau_\mu)}{\pi^2} \right\}
   \prod_{i=1}^d \rme^{2\rmi\, {\rm Im}(\gamma_i u_i - \eps_i v_i)}
\notag\\
&\qquad \alpha_i= \sum_{\mu=1}^{R-1} \sigma_\mu\; w_{i\mu}^*
\qquad \beta_i= \sum_{\mu=1}^{R-1} \sigma_\mu\; z_{i\mu}^* \qquad
\gamma_i= \sum_{\mu=1}^{R-1} \tau_\mu\; z_{i\mu} \qquad
\eps_i= \sum_{\mu=1}^{R-1} \tau_\mu\; w_{i\mu} \; .
\end{align}
Insertion into equation~(\ref{SR:eq1}), and the exchange of the order of 
integration gives:
\begin{align}
{\cal J}_4^{(R)}(\ldots) &= \frac{C_4(d,R)}
   {\Gamma\left(2(d-R+1) + (\bar m +\bar m' +\bar n' +\bar n)/2\right)}
   \prod_{\mu=1}^{R-1}\left\{\int\frac{\rmd^2(\sigma_\mu)}{\pi^2}
   \int\frac{\rmd^2(\tau_\mu)}{\pi^2}\right\} \notag\\
&\qquad\times \prod_{i=1}^d \int\rmd^2(u_i)\; u_i^{m_i}\, (u_i^*)^{n_i}\; 
   \rme^{- |u_i|^2}\; \rme^{2\rmi\, {\rm Im}[u_i (\alpha_i +\gamma_i)]}\;
   \int\rmd^2(v_i)\; v_i^{m'_i}\; (v_i^*)^{n'_i}\; \rme^{- |v_i|^2}\; 
   \rme^{2\rmi\, {\rm Im}[v_i (\beta_i -\eps_i)]} \notag\\
&= \frac{C_4(d,R)}{\Gamma(\ldots )}
   \prod_{\mu=1}^{R-1}\left\{\int\frac{\rmd^2(\sigma_\mu)}{\pi^2}
   \int\frac{\rmd^2(\tau_\mu)}{\pi^2}\right\}\;
   \prod_{i=1}^d f(m_i,n_i,\alpha_i +\gamma_i)\; 
      f(m'_i,n'_i, \beta_i -\eps_i) \; .
\label{SR:J4res}\end{align}
The function $f$ is the same which appeared in connection with the $R$-vector
integral for the unitary group, equation~(\ref{UR:J2}), and which has been 
computed in the appendix. Inserting the result, 
equation~(\ref{A4:res}) and expanding the powers of $(\alpha_i +\gamma_i)$ and
$(\beta_i -\eps_i)$, we obtain:
\begin{align}
f(m,n,\alpha +\gamma) &= \pi\, \rme^{-|\alpha +\gamma|^2} \sum_{\kappa=0}^p 
  (-)^{m-\kappa} {m\choose\kappa}\; {n\choose\kappa}\; \kappa!\; 
  \sum_{k=0}^{m-\kappa}\sum_{l=0}^{n-\kappa} 
   {m-\kappa\choose k}\; {n-\kappa\choose l}\; (\alpha^*)^k\; 
      (\gamma^*)^{m-\kappa-k}\; \alpha^l\; \gamma^{n-\kappa-l}\\ 
f(m',n', \beta -\eps) &= \pi\, \rme^{-|\beta-\eps|^2} \sum_{\kappa'=0}^{p'} 
   (-)^{m'-\kappa'} {m'\choose\kappa'}\; {n'\choose\kappa'}\; \kappa'!\;
   \sum_{k'=0}^{m'-\kappa'}\sum_{l'=0}^{n'-\kappa'} 
   {m'-\kappa'\choose k'}\; {n'-\kappa'\choose l'} \notag\\
&\qquad (\beta^*)^{k'}\; (-\eps^*)^{m'-\kappa'-k'}\; \beta^{l'}\; 
   (-\eps)^{n'-\kappa'-l'} \; ,
\end{align} 
where we have for the moment suppressed the index $i$.
In order to completely expand the product of the functions 
$f(m_i,n_i,\alpha_i +\gamma_i)$ and $f(m'_i,n'_i, \beta_i -\eps_i)$, we define 
additional vector indeces: $p_i = \min(m_i,n_i), p'_i= \min(m_i',n_i')$,
$\kappa_i$, and $\kappa_i'$. The complete expansion then reads:
\begin{align}
&\prod_{i=1}^d f(m_i,n_i,\alpha_i +\gamma_i)\;
   f(m'_i,n'_i, \beta_i -\eps_i)= \pi^{2d}\,
   \rme^{- \sum_i ( |\alpha_i +\gamma_i|^2 + |\beta_i -\eps_i|^2)} \notag\\
&\qquad\qquad\times
   \sum_{\vec\kappa = \vec o}^{\vec p} (-)^{\bar m -\bar\kappa} 
      {\vec m\choose\vec\kappa} {\vec n\choose\vec\kappa}
   \sum_{\vec\kappa' = \vec o}^{\vec p'} (-)^{\bar m' -\bar\kappa'} 
      {\vec m'\choose\vec\kappa'} {\vec n'\choose\vec\kappa'}\; 
   \prod_{i=1}^d\left\{ \kappa_i!\; \kappa'_i!\right\} \; 
\notag\\
&\qquad\qquad\times \sum_{\vec k=\vec o}^{\vec m-\vec\kappa}
   \sum_{\vec l= \vec o}^{\vec n-\vec\kappa}
   {\vec m-\vec\kappa\choose\vec k} {\vec n-\vec\kappa\choose\vec l}
   \sum_{\vec k'=\vec o}^{\vec m'-\vec\kappa'}
   \sum_{\vec l'= \vec o}^{\vec n'-\vec\kappa'} 
   {\vec m'-\vec\kappa'\choose\vec k'} {\vec n'-\vec\kappa'\choose\vec l'}
\notag\\
&\qquad\qquad\times \prod_{i=1}^d (\alpha_i^*)^{k_i}\; 
   (\gamma^*_i)^{m_i-\kappa_i-k_i}\; \alpha_i^{l_i}\; 
   \gamma_i^{n_i-\kappa_i-l_i}\; (\beta^*_i)^{k'_i}\; 
   (-\eps_i^*)^{m'_i-\kappa'_i-k'_i}\; \beta_i^{l'_i}\; 
   (-\eps_i)^{n'_i-\kappa'_i-l'_i}\; .
\end{align}
The argument of the exponential function can be simplified, as follows:
\begin{align}
&\sum_i |\alpha_i +\gamma_i|^2 + \sum_i |\beta_i -\eps_i|^2 =
\sum_{\mu\nu}\sum_i \left[ (\sigma_\mu w_{i\mu}^* +\tau_\mu z_{i\mu})
   (\sigma_\nu^* w_{i\nu} +\tau_\nu^* z_{i\nu}^*) + 
   (\sigma_\mu z_{i\mu}^* - \tau_\mu w_{i\mu})
   (\sigma_\nu^* z_{i\nu} - \tau_\nu^* w_{i\nu}^*) \right] \notag\\
&\qquad = \sum_{\mu\nu} \left[
   \sigma_\mu\sigma_\nu^* 
      (\la\vec w_\mu|\vec w_\nu\ra + \la\vec z_\mu|\vec z_\nu\ra) +
   \tau_\mu\sigma_\nu^*     
      (\la\vec z_\mu^*|w_\nu\ra - \la\vec w_\mu^*|\vec z_\nu\ra) +
   \sigma_\mu\tau_\nu^* 
      (\la\vec w_\mu|\vec z_\nu^*\ra - \la\vec z_\mu|\vec w_\nu^*\ra) +
   \tau_\mu \tau_\nu^*    
      (\la\vec z_\mu^*|\vec z_\nu^*\ra + \la\vec w_\mu^*|\vec w_\nu^*\ra) 
\right] \notag\\
&\qquad = \sum_\mu (|\sigma_\mu|^2 + |\tau_\mu|^2) \; ,
\end{align}
where we have used the orthogonality relations in~(\ref{GS:ONrels}).
For the expansion of the powers of coefficients 
$\alpha_i^*, \gamma_i^*$, etc., we will need eight matrix indices, each of 
dimension $d$$\times$$d$. It will be convenient to combine them into two
$2$$\times$$2$ block-matrices, as follows:
\begin{equation}
\bm K_1= \begin{pmatrix} L_1' & K_1\\ L_1 & K_1'\end{pmatrix}\; ,\qquad
\bm K_2= \begin{pmatrix} L_2' & K_2\\ L_2 & K_2'\end{pmatrix}\; .
\end{equation}
Then we combine pairs of coefficients, which lead to monomials of entries
from the same submatrix of $\bm w$:
\begin{align}
\prod_{i=1}^d (\alpha_i^*)^{k_i}\; (-\eps_i)^{n_i'-\kappa_i'-l_i'}
&= (-)^{\bar n' -\bar\kappa' -\bar l'}\;  
   \sum_{K_1, K_2} (\vec k|K_1)\, (\vec n' -\vec\kappa' -\vec k'|K_2)\; 
   \prod_{\mu=1}^{R-1} (\sigma_\mu^*)^{\bar k^1}\; \tau_\mu^{\bar k^2_\mu}\; 
   \prod_{i=1}^d w_{i\mu}^{K^1_{i\mu} +K^2_{i\mu}} \\
\prod_{i=1}^d (\beta_i^*)^{k_i'}\; \gamma_i^{n_i -\kappa_i -\l_i} 
&= \sum_{K_1',K_2'} (\vec k'|K_1')\, (\vec n -\vec\kappa -\vec l|K_2')\;
   \prod_{\mu=1}^{R-1} (\sigma_\mu^*)^{\bar k^{1\prime}_\mu}\; 
      \tau_\mu^{\bar k^{2\prime}_\mu}\; 
   \prod_{i=1}^d z_{i\mu}^{K^{1\prime}_{i\mu} +K^{2\prime}_{i\mu}} \\
\prod_{i=1}^d \alpha_i^{l_i}\; (-\eps_i^*)^{m'_i-\kappa'_i-k'_i} 
&= (-)^{\bar m' -\bar\kappa' -\bar k'} 
   \sum_{L_1,L_2} (\vec l|L_1)\, (\vec m' -\vec\kappa' -\vec k'|L_2)\;
   \prod_{\mu=1}^{R-1} \sigma_\mu^{\bar l^1_\mu}\; 
      (\tau_\mu^*)^{\bar l^2_\mu}\;
   \prod_{i=1}^d (w_{i\mu}^*)^{L^1_{i\mu} +L^2_{i\mu}}\\
\prod_{i=1}^d (\gamma^*_i)^{m_i-\kappa_i-k_i}\; \beta_i^{l_i'}
&= \sum_{L_1',L_2'} (\vec l'|L_1')\, (\vec m -\vec\kappa -\vec k|L_2')\;
   \prod_{\mu=1}^{R-1} \sigma_\mu^{\bar l^{1\prime}_\mu}\; 
      (\tau_\mu^*)^{\bar l^{2\prime}_\mu}\;
   \prod_{i=1}^d (z_{i\mu}^*)^{L^{1\prime}_{i\mu} +L^{2\prime}_{i\mu}} \; ,
\end{align}
where we eventually convert the subscripts $1,2$ of the eight matrix indices
into superscripts, especially when we refer to particular matrix elements.
The parameters $\bar k^1_\mu, \bar k^{1\prime}_\mu$, etc., denote the column
sums of the matrix indices $K_1, K_1'$, etc.\ . Plugging the complete expansion
into~(\ref{SR:J4res}), we obtain:
\begin{align}
&{\cal J}_4^{(R)}(\vec m,\vec m',\vec n,\vec n') = 
   \frac{C_4(d,R)\; \pi^{2d}\; (-)^{\bar m+\bar m'}}
   {\Gamma\left(2(N-R+1) +\frac{\bar m+\bar m'+\bar n+\bar n'}{2}\right)}\;
   \sum_{\vec\kappa = \vec o}^{\vec p}\sum_{\vec\kappa' = \vec o}^{\vec p'} 
   (-)^{\bar\kappa +\bar\kappa'} {\vec m\choose\vec\kappa}\; 
   {\vec m'\choose\vec\kappa'}\; {\vec n\choose\vec\kappa}\; 
   {\vec n'\choose\vec\kappa'}\; 
   \prod_{i=1}^d\left\{ \kappa_i!\; \kappa'_i!\right\}\;
\notag\\
&\qquad\qquad\times \sum_{\vec k=\vec o}^{\vec m-\vec\kappa}
   \sum_{\vec k'=\vec o}^{\vec m'-\vec\kappa'}
   \sum_{\vec l= \vec o}^{\vec n-\vec\kappa}
   \sum_{\vec l'= \vec o}^{\vec n'-\vec\kappa'} (-)^{\bar k'+\bar l'}\; 
   {\vec m-\vec\kappa\choose\vec k}\; {\vec m'-\vec\kappa'\choose\vec k'}\; 
   {\vec n-\vec\kappa\choose\vec l}\; {\vec n'-\vec\kappa'\choose\vec l'}
\notag\\
&\qquad\qquad\times
   \sum_{\bm K_1, \bm K_2} 
      (\vec k|K^1) (\vec k'|K_1') (\vec l|L_1) (\vec l'|L_1')\;\;
      (\vec n' -\vec\kappa' -\vec l'|K_2) (\vec n -\vec\kappa -\vec l|K_2')
      (\vec m' -\vec\kappa' -\vec k'|L_2) (\vec m -\vec\kappa -\vec k|L_2')
\notag\\
&\qquad\qquad\times
   \prod_{\mu=1}^{R-1}\left\{\int\frac{\rmd^2(\sigma_\mu)}{\pi^2}
   \int\frac{\rmd^2(\tau_\mu)}{\pi^2}\; 
      \sigma_\mu^{\bar l^1_\mu +\bar l^{1\prime}_\mu}\; 
      (\sigma^*_\mu)^{\bar k^1_\mu +\bar k^{1\prime}_\mu}\; 
      \tau_\mu^{\bar k^2_\mu +\bar k^{2\prime}_\mu}\; 
      (\tau^*_\mu)^{\bar l^2_\mu +\bar l^{2\prime}_\mu}\; 
      \rme^{-\sum_\mu ( |\sigma_\mu|^2 + |\tau_\mu|^2)}\;
   \right\}\notag\\
&\qquad\qquad\times
   \prod_{\mu=1}^{R-1} \prod_{i=1}^d 
      w_{i\mu}^{K^1_{i\mu} + K^2_{i\mu}} \; 
      z_{i\mu}^{K^{1\prime}_{i\mu} + K^{2\prime}_{i\mu}}\;
      (w_{i\mu}^*)^{L^1_{i\mu} + L^2_{i\mu}} \; 
      (z_{i\mu}^*)^{L^{1\prime}_{i\mu} + L^{2\prime}_{i\mu}} \; .
\label{SR:eq2}\end{align}
We can now merge each sum and its corresponding binomial from the second line 
with two sums and their corresponding multinomials from the third line:
\begin{align}
&\sum_{\vec k=\vec o}^{\vec m-\vec\kappa}
{\vec m-\vec\kappa\choose\vec k}
\sum_{K_1,L_2'} (\vec k|K^1)\, (\vec m -\vec\kappa -\vec k|L_2')
= \sum_{K^1,L_2'} (\vec m -\vec\kappa| L_2', K_1) \notag\\
&\sum_{\vec k'=\vec o}^{\vec m'-\vec\kappa'}
{\vec m'-\vec\kappa'\choose\vec k'}
   \sum_{K_1',L_2} (\vec k'|K_1')\, (\vec m' -\vec\kappa' -\vec k'|L_2)
= \sum_{K_1',L_2} (\vec m' -\vec\kappa'| L_2, K_1') \notag\\
&\sum_{\vec l =\vec o}^{\vec n -\vec\kappa} 
{\vec n -\vec\kappa\choose\vec l}
   \sum_{L_1,K_2'} (\vec l|L_1)\, (\vec n -\vec\kappa -\vec l|K_2')
= \sum_{L_1,K_2'} (\vec n -\vec\kappa|L_1, K_2') \notag\\
&\sum_{\vec l' =\vec o}^{\vec n' -\vec\kappa'}
{\vec n' -\vec\kappa'\choose\vec l'}
   \sum_{L_1',K_2} (\vec l'|L_1')\, (\vec n' -\vec\kappa' -\vec l'|K_2)
= \sum_{L_1',K_2} (\vec n' -\vec\kappa'|L_1', K_2) \; ,
\end{align}
where an extended vector-multinomial such as $(\vec m|K_1,L_2')$ means that
the rows of $K_1$ and $L_2'$ should be concatenated, to yield a product of 
multinomials, each with $2d$ elements. Evaluating the final integrals on 
$\sigma_\mu$ and $\tau_\mu$, we obtain:
\begin{align}
&{\cal J}_4^{(R)}(\vec m,\vec m',\vec n,\vec n') = 
   \frac{C_4(d,R)\; \pi^{2d-R+1}\; (-)^{\bar m+\bar m'}}
   {\Gamma\left(2(N-R+1) +\frac{\bar m+\bar m'+\bar n+\bar n'}{2}\right)}\;
   \sum_{\vec\kappa = \vec o}^{\vec p}\sum_{\vec\kappa' = \vec o}^{\vec p'} 
   (-)^{\bar\kappa +\bar\kappa'} {\vec m\choose\vec\kappa}\; 
   {\vec m'\choose\vec\kappa'}\; {\vec n\choose\vec\kappa}\; 
   {\vec n'\choose\vec\kappa'}\; 
   \prod_{i=1}^d\left\{ \kappa_i!\; \kappa'_i!\right\}\;
\notag\\
&\qquad\times \sum_{K^1,K^2} \left( \left.
   \begin{matrix} \vec m -\vec\kappa\\ \vec m' -\vec\kappa'\end{matrix}\right|
   \begin{matrix} L_2' & K_1\\ L_2 & K_1'\end{matrix}\right)\;
   \left( \left. \begin{matrix} \vec n -\vec\kappa\\ \vec n' -\vec\kappa'
   \end{matrix}\right|\begin{matrix} L_1 & K_2'\\ L_1' & K_2\end{matrix}\right)
\; (-)^{\bar K_1' +\bar L_1'}\; \prod_{\mu=1}^{R-1} 
\delta_{\bar l^1_\mu +\bar l^{1\prime}_\mu,\bar k^1_\mu + \bar k^{1\prime}_\mu}
\delta_{\bar l^2_\mu +\bar l^{2\prime}_\mu,\bar k^2_\mu + \bar k^{2\prime}_\mu}
\; (\bar k^1_\mu + \bar k^{1\prime}_\mu)!\, 
   (\bar k^2_\mu + \bar k^{2\prime}_\mu)!\notag\\
&\qquad\times
\prod_{i=1}^d 
      w_{i\mu}^{K^1_{i\mu} + K^2_{i\mu}} \;
      z_{i\mu}^{K^{1\prime}_{i\mu} + K^{2\prime}_{i\mu}} \;
      (w_{i\mu}^*)^{L^1_{i\mu} + L^2_{i\mu}} \; 
      (z_{i\mu}^*)^{L^{1\prime}_{i\mu} + l^{2\prime}_{i\mu}} \; .
\label{SR:eq3}\end{align}
where, we have replaced $\bar k'$ and $\bar l'$ with $\bar K_1'$ and 
$\bar L_1'$, as a more appropriate notation for the sum over all matrix entries
in $K_1'$ and $L_1'$, respectively. We have also further expanded the vector
multinomials, concatenating the matrices and vectors in the vertical direction.

Now, we may insert this result into equation~(\ref{SR:N}) in order to obtain
$\cN_4^{(R)}(\bm M)$ as a linear combination of terms which contain the
$(R$-$1)$-vector integrals $\cN_4^{(R-1)}(\bm M^{(R-1)} + \bm K_1 +\bm K_2)$. 
Note that the matrix indices
in~(\ref{SR:eq3}) are chosen in order to fit into that scheme. Dividing through
the corresponding expressions for $\cN_4^{(R)}(\bm o)$ and 
$\cN_4^{(R-1)}(\bm o)$ we obtain:
\begin{align}
\la\bm M\ra &= \frac{\Gamma\left( 2(d-R+1)\right)\, (-)^{\bar m +\bar m'}}
   {\Gamma\left( 2(d-R+1) + (\bar m +\bar m' +\bar n +\bar n')/2\right)}
   \sum_{\vec\kappa, \vec\kappa' = \vec o}^{\vec p, \vec p'} 
      (-)^{\bar\kappa + \bar\kappa'}\;
   {\vec m\choose\vec\kappa}\; {\vec m'\choose\vec\kappa'}\; 
   {\vec n\choose\vec\kappa}\; {\vec n'\choose\vec\kappa'}\; 
   (2d)_{\bar\kappa +\bar\kappa'}\;
   \lla\left.\begin{matrix} \vec\kappa'\\ \vec\kappa\end{matrix}\right|
      \begin{matrix} \vec\kappa\\ \vec\kappa'\end{matrix}\rra \notag\\
&\qquad\times \sum_{K^1,K^2} \left( \left.
   \begin{matrix} \vec m -\vec\kappa\\ \vec m' -\vec\kappa'\end{matrix}\right|
   \begin{matrix} L_2' & K_1\\ L_2 & K_1'\end{matrix}\right)\;
   \left( \left. \begin{matrix} \vec n -\vec\kappa\\ \vec n' -\vec\kappa'
   \end{matrix}\right|\begin{matrix} L_1 & K_2'\\ L_1' & K_2\end{matrix}\right)
\; (-)^{\bar K_1' +\bar L_1'}\; 
   (2d)_{\bar m -\bar\kappa + \bar m' -\bar\kappa'}\;
   \lla\left. \begin{matrix} \vec l^2_{\rm cs}\\
      \vec l^1_{\rm cs}\end{matrix}\right|\begin{matrix} \vec k^1_{\rm cs}\\
      \vec k^2_{\rm cs}\end{matrix}\rra \notag\\
&\qquad\times \la\bm M^{(R-1)} + \bm K_1 + \bm K_2\ra \; .
\end{align}
The vectors 
$\vec k^1_{\rm cs},\, \vec k^2_{\rm cs},\, \vec l^1_{\rm cs}$, and
$\vec l^2_{\rm cs}$ are $d$-dimensional vectors which contain in their first
$R-1$ components the column sums of $\bm K_1$ and $\bm K_2$ in the following
order: $\vec k^1_{\rm cs}$ contains the column sums of $K_1$ and $K_1'$;
$\vec k^2_{\rm cs}$ contains those of $K_2$ and $K_2'$; 
$\vec l^1_{\rm cs}$ those of $L_1$ and $L_1'$; and
$\vec l^2_{\rm cs}$ those of $L_2$ and $L_2'$.
Note that due to the one vector integral in these column sums, one may 
conclude that 
\begin{equation}
\sum_{\mu=1}^{R-1} \bar k^1_\mu + \bar k^{1\prime}_\mu + \bar k^2_\mu  
   + \bar k^{2\prime}_\mu = \bar l^1_\mu + \bar l^{1\prime}_\mu + \bar l^2_\mu
   + \bar l^{2\prime}_\mu \quad\Rightarrow\quad 
\bar m -\bar\kappa + \bar m' -\bar\kappa' = \bar n -\bar\kappa + \bar n' 
   -\bar\kappa' \quad\Rightarrow\quad \bar m + \bar m' = \bar n + \bar n' \; .
\label{S:vcond}\end{equation}
This allows us to use again the function $B(a,b;z_1,z_2)$ defined in 
equation~(\ref{O:defB}):
\begin{align}
\la\bm M\ra &= \delta_{\bar m +\bar m', \bar n +\bar n'} 
   \sum_{\vec\kappa, \vec\kappa' = \vec o}^{\vec p, \vec p'} 
   {\vec m\choose\vec\kappa}\; {\vec m'\choose\vec\kappa'}\; 
   {\vec n\choose\vec\kappa}\; {\vec n'\choose\vec\kappa'}\; 
   B\left(\bar m +\bar m', \bar\kappa +\bar\kappa'; 2d, 2(R-1)\right)\;
   \lla\left.\begin{matrix} \vec\kappa'\\ \vec\kappa\end{matrix}\right|
      \begin{matrix} \vec\kappa\\ \vec\kappa'\end{matrix}\rra \notag\\
&\qquad\times \sum_{\bm K_1, \bm K_2} \left(\left. 
   \begin{matrix} \vec m -\vec\kappa\\ \vec m' -\vec\kappa'\end{matrix}\right|
   \begin{matrix} L_2' & K_1\\ L_2 & K_1'\end{matrix}\right) \left(\left.
   \begin{matrix} \vec n -\vec\kappa\\ \vec n' -\vec\kappa'\end{matrix}\right|
   \begin{matrix} L_1 & K_2'\\ L_1' & K_2\end{matrix}\right)\;
    (-)^{\bar K_1' +\bar L_1'}\; 
   \lla\left. \begin{matrix} \vec l^2_{\rm cs}\\
      \vec l^1_{\rm cs}\end{matrix}\right|\begin{matrix} \vec k^1_{\rm cs}\\
      \vec k^2_{\rm cs}\end{matrix}\rra \;
   \la\bm M^{(R-1)} + \bm K_1 + \bm K_2\ra \; .
\label{SR:res}\end{align}
The first line of this expression agrees precisely with the corresponding
part of the recurrence relation for the unitary group $U(2d)$. The differences
appear in the second line, where we have to sum over 2$\times$2-block matrices 
$\bm K_1$ and $\bm K_2$, with the restriction that the row-sums of the 
block-matrices in the following vector-multinomials must agree with the
corresponding component of the given $2d$-dimensional vector.

\paragraph{Zero column vectors}
In the case that the $R$'th column vector of all four matrices $M$, $M'$, $N$,
and $N'$ contain only zeros, the summation over $\vec\kappa$ and $\vec\kappa'$ 
contains only one term, namely $\vec\kappa= \vec\kappa'= \vec o$. Similarly,
the summation over $\bm K_1, \bm K_2$ also contains only one term,
where all matrix entries of $\bm K_1$ and $\bm K_2$ are zero. Therefore, we
find
\begin{equation}
\la\bm M\ra= \lla\begin{pmatrix} N' & M\\ N & M'\end{pmatrix}\rra =
\lla\begin{pmatrix} N^{\prime(R-1)} & M^{(R-1)}\\ N^{(R-1)} & M^{\prime(R-1)}
\end{pmatrix}\rra = \la\bm M^{(R-1)}\ra \; .
\end{equation}
As in the case of $O(d)$ and $U(d)$ in the previous sections, this justifies 
equation~(\ref{S:cN}) where we have ignored column vectors of $w$ and $z$ 
whose entries do not occur in $\cM_4(\bm w)$.

\paragraph{Vanishing integrals}
The monomial integral $\la\bm M\ra$ obviously vanishes if $\bar m +\bar m'$ is 
different from $\bar n+ \bar n'$. Taking into account the transformations
from~(\ref{S:inv}), which do not change the integral, this means that the
integral vanishes unless the column sums of $\bm M$ fulfill the condition:
\begin{equation}
\forall\; 1\le j\le d \; :\; \bar{\bm m}_j = \bar{\bm m}_{d+j} \; ,
\end{equation}
where $\bar{\bm m}_j$ is the column sum of column $j$ of $\bm M$. An exactly 
analogous condition holds for the row sums.

\section{\label{C} Conclusions}

We applied the {\em column vector method}, originally developed for
the orthogonal group in~\cite{Gor02}, to monomial integrals over the unitary
and the unitary symplectic group. As in the orthogonal case, we obtained
recursive integration formulas, where the recursion parameter is
the number of non-empty column vectors (these are those matrix columns of the
group element, which contain at least one matrix element from the monomial).
In all three cases, the recursion formulas are of a very similar form and
relatively easy to implement in a computer algebra environment. This provides
an efficient way to compute such integrals analytically as a function of the
matrix dimension.

The approach presented here is very different from the group theoretical 
approach developped in~\cite{ColSni06}, which also provides explicite formulas 
for the analytical calculation of arbitrary monomial integrals over the 
classical groups. We did not yet compare the efficiency of both methods, but we 
would expect that the result of such a comparison will depend on the problem at 
hand.


\begin{appendix}

\section{\label{A4} Gaussian integral on the complex plane}

Here, we compute the Gaussian integral from equation~(\ref{UR:J2}). Writing
for the complex integration variable $u= x+\rmi y$ we get:
\begin{equation}
f(m,n,\alpha) = \iint\rmd x\, \rmd y\; u^m\; (u^*)^n\; \rme^{-u^* u}\;
   \rme^{\alpha\, u - \alpha^* u^*} \; .
\label{A4:defint}\end{equation}
First note that $f(m,n,\alpha)^* = f(n,m,-\alpha)$, which implies that it is 
sufficient to consider the case $m\ge n$. Then, we use the complex differential 
operators $\partial_z$ and $\bar\partial_z$ to write:
\begin{equation}
f(m,n,\alpha) = (-)^n\; \partial_\alpha^m\; \bar\partial_\alpha^n\;
   \iint\rmd x\, \rmd y\; \rme^{-u^* u}\; \rme^{\alpha\, u - \alpha^* u^*} 
\qquad
2\, \partial_z = \partial_x -\rmi\partial_y \qquad 
2\, \bar\partial_z = \partial_x +\rmi\partial_y \; .
\end{equation}
With $\alpha= \alpha_1 +\rmi\alpha_2$, it follows
\begin{align}
f(m,n,\alpha) &= (-)^n\; \partial_\alpha^m\; \bar\partial_\alpha^n\;
   \int\rmd x\; \rme^{-(x^2-2\rmi \alpha_2\, x)}\; 
   \int\rmd y\; \rme^{-(y^2-2\rmi \alpha_1\, y)} \notag\\
&= (-)^n\; \partial_\alpha^m\; \bar\partial_\alpha^n\; \pi\; 
   \rme^{-\alpha^*\,\alpha} 
 = \pi\; (-)^{m+n}\; \bar\partial_\alpha^n\; (\alpha^*)^m\; 
      \rme^{-\alpha^*\,\alpha} 
 = \pi\; (-)^{m+n}\; P(m,n)\; \rme^{-\alpha^*\,\alpha}\; ,
\end{align}
where $P(m,n)$ is a polynomial of total order $m+n$, with the property
that the difference between the power of the conjugated variable and that
of the non-conjugated one is $m-n$, fixed. Hence, for $P(m,n)$ we consider 
the following Ansatz:
\begin{equation}
P(m,n)= \sum_{k=0}^n C^{(m,n)}_k\; (\alpha^*)^{m-n+k}\; \alpha^k 
 = (a^*)^{m-n}\sum_{k=0}^n C^{(m,n)}_k\; (\alpha^*\,\alpha)^k \; .
\end{equation}
For $n=0$, the polynomials $P(m,0)$ are known, while for $m\ge n >0$, they
can be computed by recursion:
\begin{align}
n=0 &:\quad P(m,0) = 1 \quad\Rightarrow\quad C^{(m,0)}_0 = 1 \notag\\
m>0 &:\quad P(m,n) = \rme^{\alpha^*\,\alpha}\; \bar\partial_\alpha^n\; 
   (\alpha^*)^m\; \rme^{-\alpha^*\,\alpha} \notag\\
&\;\quad P(m,n+1) = \rme^{\alpha^*\,\alpha}\; \bar\partial_\alpha^n\;
   P(m,n)\; \rme^{-\alpha^*\,\alpha}
   = -\alpha\; P(m,n) + \sum_{k=0}^n C^{(m,n)}_k\, (m-n+k)\; 
      (\alpha^*)^{m-(n+1)+k}\; \alpha^k \; .
\end{align}
Equating the coefficients for corresponding powers, we obtain for 
$m\le n\le k\le 0$:
\begin{equation}
C_k^{(m,n+1)} = \begin{cases} (m-n)\; C_0^{(m,n)} &: k=0\\
   (m-n+k)\; C_k^{(m,n)} - C_{k-1}^{(m,n)} &: 1\le k\le n\\
   -C_n^{(m,n)} &: k=n+1 \end{cases} \qquad C_0^{(m,0)} = 1\; .
\end{equation}
Ordered according to $n$ and $k$, these coefficients can be arranged in a 
two-dimensional scheme, forming a Pascal triangle. From this scheme it is
easily seen that
\begin{equation}
C_k^{(m,n)} = (-)^k\; (m-n+k+1)\, (m-n+k+2)\, \ldots\, m\; {n\choose k} \; .
\end{equation}
Therefore, we finally obtain
\begin{equation}
P(m,n)= \sum_{k=0}^n (-)^k \frac{m!\; n!}{(m-n+k)!\, k!\, (n-k)!}\; 
   (\alpha^*)^{m-n+k}\; \alpha^k
 = \sum_{k=0}^n (-)^{n-k}\; \frac{m!\; n!}{(m-k)!\, k!\, (n-k)!}\;
   (\alpha^*)^{m-k}\; \alpha^{n-k} \; .
\end{equation}
As mentioned before, here we have assumed that $m\ge n$. For $m<n$ by contrast, 
we use: $f(m,n,\alpha)= f^*(n,m,-\alpha)$ so that in summary, we may write:
\begin{equation}
f(m,n,\alpha)= \pi\; (-)^m \sum_{k=0}^p (-)^k\; k!\; {m \choose k}\; 
   {n \choose k}\; (\alpha^*)^{m-k}\; \alpha^{n-k}\; \rme^{-\alpha^*\alpha} 
\qquad p= \min(m,n)\; .
\label{A4:res}\end{equation}

\end{appendix}

\bibliography{/home/gorin/Bib/amol,/home/gorin/Bib/rom,/home/gorin/Bib/deco,/home/gorin/Bib/ranh,/home/gorin/Bib/semic,/home/gorin/Bib/stas,/home/gorin/Bib/books}

\end{document}